\documentclass{jfm}
\usepackage{graphicx}  
\usepackage{esvect}
\usepackage{epstopdf,epsfig}
\usepackage{textgreek}
\usepackage[scr=bickham]{mathalfa}
\usepackage{amsmath}
\numberwithin{equation}{section}
\usepackage{bm}
\usepackage{subfigure}\usepackage{float}\usepackage{epsfig}
\usepackage{amssymb}\usepackage{psfrag}\usepackage{natbib}
\usepackage{stmaryrd}
\usepackage{picture}
\usepackage{float}
\usepackage[dvipsnames]{xcolor}
\usepackage[T1]{fontenc}
\usepackage{multirow}
\usepackage{overpic}
\usepackage{tikz}
\usepackage{subcaption}\usepackage{float}\usepackage{epsfig}
\DeclareCaptionLabelFormat{new}{\textup{(}{\text{#2}}\textup{)}}
\usepackage{stmaryrd}
\usepackage{float}
\usepackage{bm}
\usepackage{multirow}
\usepackage{makecell}
\usepackage{ragged2e}

\newcommand{\be}{\begin{equation}}
\newcommand{\ee}{\end{equation}}
\newcommand{\bef}{\begin{figure}}
\newcommand{\ef}{\end{figure}}
\newcommand{\bear}{\begin{eqnarray}}
\newcommand{\ear}{\end{eqnarray}}
\newcommand{\barr}{\begin{array}}
\newcommand{\earr}{\end{array}}

\newcommand{\bs}{\boldsymbol}

\newcommand{\msb}{\mathbf}

\definecolor{bittersweet}{rgb}{1.0, 0.44, 0.37}
\definecolor{bluep}{rgb}{0.2, 0.2, 0.6}
\definecolor{bole}{rgb}{0.47, 0.27, 0.23}
\definecolor{cgreen}{rgb}{0.0, 0.42, 0.24}
\definecolor{dpurple}{rgb}{0.59, 0.44, 0.84}

\usepackage[dvipsnames]{xcolor}
\usepackage{ragged2e}
\usepackage{physics}
\usepackage{upgreek}
\usepackage[colorlinks=true,breaklinks=true,linkcolor=blue]{hyperref} 
\shortauthor{Arun Krishna B. J. and Ganesh Tamadapu}
\title{Finite amplitude shape oscillations of magnetic microbubble under uniform magnetic field}

\author{Arun Krishna B. J.\aff{} \and Ganesh Tamadapu\aff{}
  \corresp{\email{gt@iitm.ac.in}}}

\affiliation{
  Department of Applied Mechanics and Biomedical Engineering, Indian Institute of Technology Madras,
Chennai, 600036, India.}
\begin{document}
\maketitle
\begin{abstract}
This study presents a theoretical model for the non-spherical oscillations of a lipid-encapsulated bubble infused with magnetic particles and subjected to a uniform magnetic field. Since the applied magnetic field is uniform, its gradient vanishes, making the existing weakly magnetic membrane theory inapplicable. To overcome this limitation, an alternative formulation that accounts for both the thinness and weak magnetic nature of the bubble shell is developed. Using this formulation, the induced magnetic field is calculated for a spherical bubble and incorporated into a Lagrangian framework for describing the non-spherical oscillations. The resulting governing equations for the radial and shape modes are then derived. The analysis shows that both the natural frequencies and exponential stability of the shape modes increase with increasing applied magnetic field. The frequency response of the shape modes shifts towards higher frequencies as the applied magnetic field is increased, consistent with the increase in their natural frequencies. Furthermore, finite-amplitude oscillations exhibit a period-doubling sequence and progressively transition towards chaotic behavior with increasing magnetic field strength. 
\end{abstract}

\maketitle
\section{Introduction}\label{s:1}
In recent years, targeted delivery strategies have shown significant promise in enhancing the efficacy of diverse pharmaceutical agents while minimizing adverse side effects. As a result, they are emerging as a key component of next-generation therapies with improved safety profiles. Magnetic microbubbles (MMBs) are microbubbles (MBs) coated with magnetic nanoparticles (NPs). In addition to preserving the acoustic properties of conventional MBs, MMBs can also function as effective contrast agents for magnetic resonance imaging (MRI)~\cite{stride2009enhancement,sciallero2016magnetic,chertok2018circulating}. This dual-modality capability makes MMBs highly attractive for a broad range of biomedical applications, including targeted and localized drug/gene delivery. Adding magnetic nanoparticles to encapsulation has three key additional features making them attractive for biomedical applications: their ability to be visualized under MRI \cite{sun2008magnetic}, to be localized at a target site by means of a magnetic field gradient \cite{pankhurst2003applications}, and to produce localized heating when exposed to an oscillating magnetic field for hyperthermia and/or to promote drug delivery \cite{duran2008magnetic,thiesen2008clinical}. Furthermore,
\cite{duan2015controlled} reported that increasing the concentration of magnetic nanoparticles attached to the microbubbles enhanced MRI visibility while also prolonging the persistence of the ultrasound signal. \cite{owen2015magnetic} treated magnetic bubbles as rigid particles and examined their dynamics in Poiseuille flow and investigated the magnetic field conditions needed to retain them near the vessel wall. However, developing magnetic microbubbles that can simultaneously fulfill diagnostic and therapeutic objectives remains a major challenge. In particular, there remains insufficient understanding of how incorporating magnetic nanoparticles into microbubble shells alters their mechanical response and dynamics in ultrasound-assisted imaging and drug delivery.

Addressing these difficulties is crucial for the development of effective magnetic microbubbles. Modeling the oscillatory behavior of encapsulated magnetic microbubbles poses significant theoretical challenges because the absence of magnetic monopoles leads to inherently axisymmetric oscillations. As a result, non-spherical modes arise directly from the imposed magnetic field rather than emerging as parametric instabilities of the spherical mode. The nonspherical dynamics of gas bubbles have been extensively studied through both theoretical and experimental investigations in~\cite{hao1999effect,guedra2016experimental,guedra2018bubble,shaw2006translation,shaw2009stability,shaw2017nonspherical}. Numerical studies by ~\cite{shaw2006translation} and experimental observations by \cite{guedra2016experimental} demonstrated that nonlinear mode coupling plays a key role in saturating the instabilities.

Studies addressing the non-spherical oscillations of encapsulated bubbles remain relatively limited. \cite{tsiglifis2011parametric} investigated the parametric stability and dynamic buckling behavior of encapsulated bubbles, whereas \cite{liu2012surface} examined their surface instabilities and analyzed the influence of membrane surface properties on the natural frequencies of the bubble shell. Their results further indicated that the parametric resonance characteristics of encapsulated bubbles are similar to those of free gas bubbles, with membrane effects exerting little influence on stability. However, these models are limited by their assumption of linear shape-mode oscillations. To overcome this limitation, \cite{dash2024nonspherical} incorporated nonlinear mode coupling and interfacial energy effects into the model, demonstrating that nonlinearities and interface effects also play a crucial role in saturating instabilities in encapsulated bubbles.

Few works deal with modeling magnetic microbubbles. The study by \cite{zhao2022nonlinear} focused on encapsulated bubbles undergoing radial oscillations in an applied magnetic field, while \cite{du2024local} further accounted for translational dynamics in the $x$ and $y$ directions in addition to radial motion. Recently, \cite{bj2026non} developed equations for non-spherical oscillations of bubbles placed under current-carrying coils and a magnetic dipole using the theory developed by \cite{barham2007finite,barham2012magnetoelasticity}, where forcing only due to the applied field is considered. Although the model in~\cite{bj2026non} successfully describes oscillations in non-uniform magnetic fields, it cannot capture the dynamics under a uniform magnetic field, as the vanishing field gradient removes the magnetic forcing effect. 

The uniformity of the magnetic field is essential in MRI systems for achieving proper imaging performance and high image quality~\cite{moser2009magnetic}. Consequently, modeling bubble oscillations under a uniform magnetic field is also of considerable importance. However, under a uniform field, the gradient of the applied magnetic field vanishes, making it necessary to evaluate the magnetic field induced within the membrane and its associated gradient in order to determine the magnetic forcing. In the present work, a simplified formulation for calculating the induced magnetic field is derived following the approaches of ~\cite{gu2011coordinate,gu2011interfacial}. These expressions are then used to compute the magnetic field distribution and the resulting magnetic forces. The energy due to this magnetic field is added to the membrane energy, and this is introduced in the Lagrangian energy density expression given in~\cite{shaw2006translation}. Using this, the governing equations for radial and shape modes are obtained, considering only radial shape modes. Only quadratic interaction between shape modes is considered in the analysis. Using these equations, the influence of the magnetic field on bubble oscillations is analyzed.

The paper is organized as follows. Section \ref{s:2} provides the mathematical formulation of the problem, including the modified theory to calculate the induced magnetic field, kinematics of the bubble surface, and Lagrangian formulation for mode equations. In Section \ref{s:3}, the influence of the static magnetic field on natural frequency, stability diagram, frequency response, and finite-amplitude time-series oscillations is discussed. Finally, Section \ref{s:4} offers a summary and a conclusion.

\section{Mathematical formulation}\label{s:2}
\subsection{Theory for magnetic field induced in thin magnetic interfaces}
We consider a gas filled magnetic microbubble placed in a fluid under uniform magnetic field and acoustic pressure. In general, analytically calculating the induced magnetic field for a finitely deforming three-dimensional body is highly challenging. Therefore, a simplified theory for evaluating the induced field in thin membranes is developed in this section. Later, this theory is applied to the thin microbubble encapsulation to calculate induced field in a deforming bubble.  
\subsubsection{Non-deformable interface}
\begin{figure}
    \centering
    \includegraphics[width=0.8\linewidth]{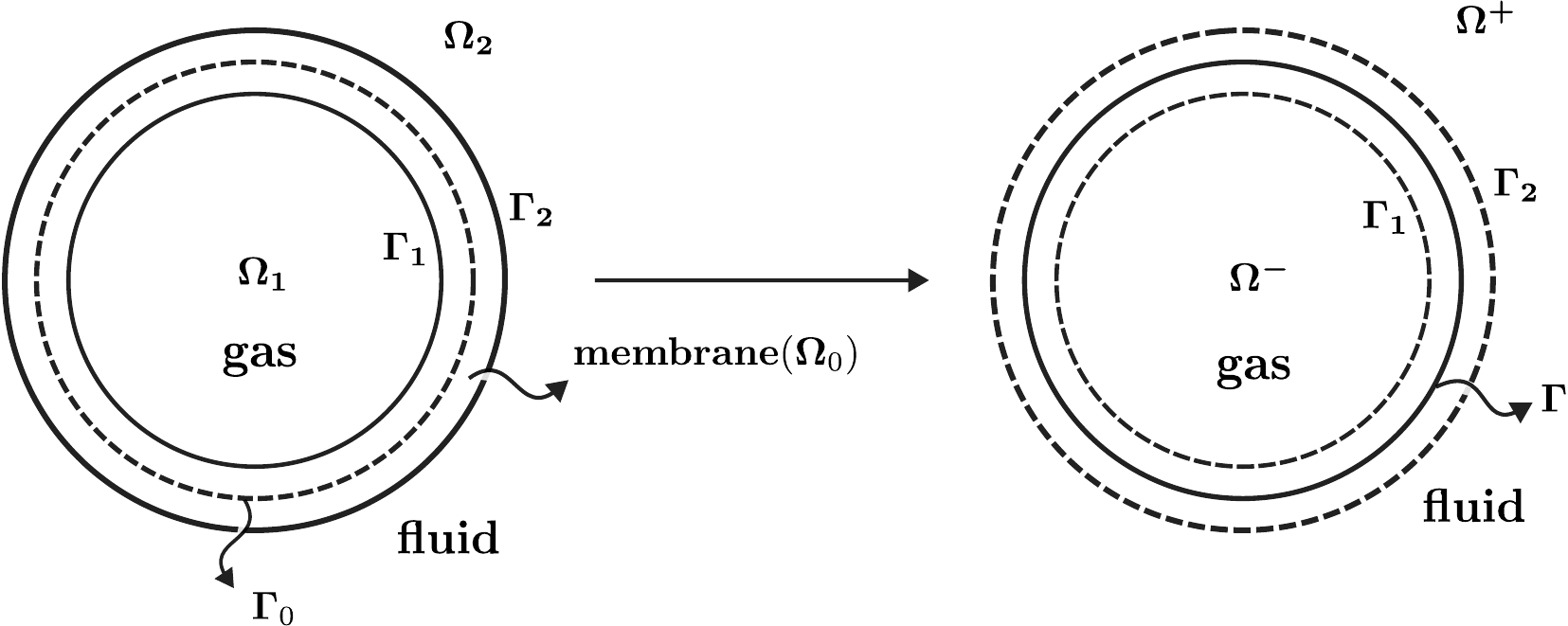}
    \caption{ Three phase configuration (left) gas $(\Upomega_1)$- membrane $(\Upomega_0)$- fluid $(\Upomega_2)$ replaced by equivalent  two phase configuration (right) of gas $(\Upomega^-)$- interface $(\Upgamma)$- fluid $(\Upomega^+)$.}
    \label{fig:1}
\end{figure}

The theory for calculating the induced magnetic field is first derived for a rigid thin membrane and subsequently generalized for deformable interfaces.  The three-dimensional Maxwell equations for the magnetic field are,
\begin{subequations}\label{e:1}
\begin{align}
\bs\nabla\cdot \bs{b}&=\mathbf{0},\label{e:1_a}\\
\bs\nabla\times\bs{h}&=\mathbf{0},\label{e:1_b}
\end{align}
\end{subequations}
where $\bs{b}$ is the magnetic flux whose normal component is continuous across a surface, and $\bs{h}$ is the magnetic field whose tangential component is continuous across a surface. 
\noindent As the thickness $(T)$ of the membrane is minimal compared to its other dimensions, the three-phase configuration is replaced by an equivalent two-phase configuration, as shown in figure~\ref{fig:1}. For this replacement to be physically consistent, the jumps in the potential and its gradient across interface $\Upgamma_1$ and $\Upgamma_2$ in both configurations are required to agree up to terms of order $\mathcal{O}(T^2)$, as adopted in~\cite{gu2011coordinate, gu2011interfacial}. Let $\phi^0$, $\phi^1$ and $\phi^2$ be potential in membrane ${(\Upomega}_0)$, gas $({\Upomega}_1)$ and fluid $({\Upomega}_2)$ regions, respectively, as shown in figure~\ref{fig:1}. Expanding the potential inside the membrane along the normal direction, the potential at the mid-surface $(\Upgamma_0)$ is,
\begin{align}\label{e:2}
\begin{split}
\phi^0\vert_{\Upgamma_0}&=\phi^0\vert_{\Upgamma_1}+\frac{T}{2}\left(\nabla_n\phi^0\right)\vert_{\Upgamma_1}+\mathcal{O}(T^2),\\
\phi^0\vert_{\Upgamma_0}&=\phi^0\vert_{\Upgamma_2}-\frac{T}{2}\left(\nabla_n\phi^0\right)\vert_{\Upgamma_2}+\mathcal{O}(T^2).
\end{split}
\end{align}
Using continuity of potential $(\phi^i\vert_{\Upgamma_i}=\phi^0\vert_{\Upgamma_i})$ at interfaces $\Upgamma_1$ and $\Upgamma_2$, the jump in potential along membrane thickness is expressed as
\begin{align}\label{e:3}
\phi^2\vert_{\Upgamma_2}-\phi^1\vert_{\Upgamma_1}&=\frac{T}{2}(\left(\nabla_n\phi^0\right)\vert_{\Upgamma_2}+\left(\nabla_n\phi^0\right)\vert_{\Upgamma_1})+\mathcal{O}(T^2).
\end{align}
where $\nabla_n=\bs{\nabla\cdot {\bs n}}$, $\bs{\nabla}$ is the gradient operator and ${\bs n}$ is the normal at mid-surface. After replacing the three-phase configuration with a two-phase configuration $(\Upomega^-)$- interphase $(\Upgamma)$- fluid $(\Upomega^+)$ (see right part of figure~\ref{fig:1}), the potential difference across $\Upgamma_2$ and $\Upgamma_1$ is,
\begin{subequations}\label{e:4}
\begin{align}\label{e:4}
\phi^+\vert_{\Upgamma_2}&=\phi^+\vert_{\Upgamma}+\frac{T}{2}\left(\nabla_n\phi^+\right)\vert_\Upgamma+\mathcal{O}(T^2)\\
\phi^+\vert_{\Upgamma_1}&=\phi^-\vert_{\Upgamma}-\frac{T}{2}\left(\nabla_n\phi^-\right)\vert_\Upgamma+\mathcal{O}(T^2)\\
\phi^+\vert_{\Upgamma_2}-\phi^-\vert_{\Upgamma_1}&=\phi^+\vert_\Upgamma-\phi^-\vert_\Upgamma+\frac{T}{2}(\left(\nabla_n\phi^+\right)\vert_\Upgamma+\left(\nabla_n\phi^-\right)\vert_\Upgamma)+\mathcal{O}(T^2).\label{e:4-c}
\end{align}
\end{subequations}
Here, $\phi^+$ and $\phi^-$ are potential in replaced gas region $\Upomega^-$ and replaced fluid region $\Upomega^+$ as shown in figure~\ref{fig:1}. Similarly, the normal component of magnetic flux at interfaces $\Upgamma_1$ and $\Upgamma_2$ in the two-phase configuration is,
\begin{align}\label{e:5}
\begin{split}
B_n^2\vert_{\Upgamma_2}&=B_n^+\vert_{\Upgamma}+\mathcal{O}(T),\\
B_n^1\vert_{\Upgamma_1}&=B_n^-\vert_{\Upgamma}+\mathcal{O}(T).
\end{split}
\end{align}
The gradient of the potential in the normal direction can be expressed in terms of normal magnetic flux $(B_n)$, membrane permeability $\mu_m$, and free space permeability $\mu_0$ as,
\begin{align}\label{e:6}
\begin{split}
\left(\nabla_n\phi^+\right)\vert_{\Upgamma}&=-\frac{B^+_n\vert_{\Upgamma}}{\mu_0},\\
\left(\nabla_n\phi^-\right)\vert_{\Upgamma}&=-\frac{B^-_n\vert_{\Upgamma}}{\mu_0},\\
\left(\nabla_n\phi^0\right)\vert_{\Upgamma_2}&=-\frac{B^2_n\vert_{\Upgamma_2}}{\mu_m}=-\frac{B^+_n\vert_{\Upgamma}}{\mu_m}+\mathcal{O}(T),\\
\left(\nabla_n\phi^0\right)\vert_{\Upgamma_1}&=-\frac{B^1_n\vert_{\Upgamma_1}}{\mu_m}=-\frac{B^-_n\vert_{\Upgamma}}{\mu_m}+\mathcal{O}(T).
\end{split}
\end{align}
Substituting these in \eqref{e:4-c}, the jump in potential across the interface in the two-phase configuration is obtained as,
\begin{align}\label{e:7}
\left[\phi\right]=\phi^+-\phi^-=\frac{T}{2}\left(\frac{1}{\mu_0}-\frac{1}{\mu_m}\right)\left(B^+_n\vert_{\Upgamma}+B^-_n\vert_{\Upgamma}\right)+\mathcal{O}(T^2).
\end{align}
Doing the same for the normal component of the magnetic field $B_n$ in a two-phase configuration, the jump in magnetic flux across the interface $\Upgamma_1$ and $\Upgamma_2$ becomes,
\begin{align}\label{e:8}
\begin{split}
B^2_n\vert_{\Upgamma_2}-B^1_n\vert_{\Upgamma_1}&=\begin{bmatrix}
B_n    
\end{bmatrix}+\frac{T}{2}\left(\nabla_nB^+_n\vert_{\Upgamma}+\nabla_nB^-_n\vert_{\Upgamma}\right)+\mathcal{O}(T^2).\\
\end{split}
\end{align}
The gradient of the normal component of magnetic flux in the normal direction can be expressed as,
\begin{align}\label{e:9}
\begin{split}
\nabla_nB_n&=\bs{\nabla}B_n\cdot\bs{n}=\bs{\nabla B}\cdot\bs{n\otimes n}=-\bs{\nabla}_s\cdot\bs{B}\\
&=\mu\bs{\nabla}_s\cdot\bs{\nabla}_s\phi-\bs{\nabla}_s\cdot(B_n\bs{n}).
\end{split}
\end{align}
where $\bs{\nabla}_s=\bs{\nabla-(\nabla\cdot n)n}$ is surface del operator. The above equation is of a general form and can be used at any interface. In a three-phase configuration, the jump in magnetic flux across $\Upgamma_1$ and $\Upgamma_2$ becomes,
\begin{align}\label{e:10}
\begin{split}
B^2_n\vert_{\Upgamma_2}-B^1_n\vert_{\Upgamma_1}&=\frac{T}{2}\left(\nabla_nB^2_n+\nabla_nB^1_n\right)\\&=\frac{T}{2}(\mu_m\bs{\nabla}_s\cdot\bs{\nabla}_s(\phi^+\vert_\Upgamma+\phi^-\vert_\Upgamma)-\bs{\nabla}_s\cdot((B_n^+\vert_\Upgamma+B_n^-\vert_\Upgamma)\bs{n})+\mathcal{O}(T^2),
\end{split}
\end{align}
Substituting~\eqref{e:5}, \eqref{e:9} and~\eqref{e:10} into equation~\eqref{e:8} the jump in the normal component of magnetic flux in the two-phase configuration becomes,
\begin{align}\label{e:11}
\left[B_n\right]=B_n^+\vert_{\Upgamma}-B^-_n\vert_{\Upgamma}=\frac{T}{2}\left(\mu_m-\mu_0\right)\left(\bs{\nabla}_s\cdot(\bs{\nabla}_s(\phi^+\vert_\Upgamma+\phi^-\vert_\Upgamma)\right)+\mathcal{O}(T^2)
\end{align}

\subsubsection{Deformable interface}
Let $\bs{X}$ be the position of a point in the reference configuration (undeformed bubble), which gets mapped to the position $\bs{x}$ in the current configuration (deformed bubble). The deformation gradient $\msb{F}$ is given by 
\begin{align}\label{e:12}
\msb{F}=\frac{\partial\bs{x}}{\partial\bs{X}}.
\end{align}
The relation between area vector $({\rm d}\bs{a},{\rm d}\bs{A})$ and length vector $({\rm d}\bs{l},{\rm d}\bs{L})$ in the current configuration and reference configuration is,
\begin{align}\label{e:13}
\begin{split}
{\rm d}\bs{a}&=J\msb{F}^{-T}\cdot {\rm d}\bs{A}\\
{\rm d}\bs{l}&=\msb{F}\cdot {\rm d}\bs{L}.
\end{split}
\end{align}
where $J=\rm{det}(\msb{F})$. Integrating governing equation~\eqref{e:1_a} over current volume ${\rm d}v$ and~\eqref{e:1_b} over current area ${\rm d}\bs{a}$, 
\begin{align}
\begin{split}
&\int_{v}\bs{\nabla}\cdot\bs{b}\,{\rm d}v=\int_{a}\bs{b}\cdot {\rm d}\bs{a}=\int_A J\msb{F}^{-1}\bs{b}\cdot {\rm d}\bs{A}=\int_{V}\bs{\nabla}_r\cdot\bs{B}\,{\rm d}V=0,\\
&\int_a\bs{\nabla}\times\bs{h}\cdot {\rm d}\bs{a}=\int_l\bs{h}\cdot {\rm d}\bs{l}=\int_L\msb{F}^{T}\bs{h}\cdot {\rm d}\bs{L}=\int_A\bs{\nabla}_r\times\bs{H}\cdot {\rm d}\bs{A}=0,
\end{split}
\end{align}
where $\bs{\nabla}_r$ is gradient operator in reference configuration. This gives pull back of magnetic field $\bs{h}$ and flux $\bs{b}$ as
\begin{subequations}\label{e:15}
\begin{align}
\bs{B}&=J\msb{F}^{-1}\bs{b}, \label{e:15_a}\\
\bs{H}&=\msb{F}^{T}\bs{h},\label{e:15_b}\\
\bs{B}&=\mu\,J\msb{C}^{-1}\bs{H},\label{e:15_c}
\end{align}
\end{subequations}
where $\msb{C}=\msb{F}^{T}\msb{F}$ is the right Cauchy-Green defomation tensor. In our study, only incompressible membranes are considered, for which the thickness is constant only in the reference configuration, in which a jump in potential and magnetic flux can be given as,
\begin{align}\label{e:16}
\begin{split}
\left[\phi\right]&=\phi^+\vert_\Upgamma-\phi^-\vert_\Upgamma=\frac{T}{2}\left(1-\frac{\mu_0}{\mu_m}\right)\left(H^+_n\vert_\Upgamma+H^-_n\vert_\Upgamma\right)+\mathcal{O}(T^2),\\
\left[B_n\right]&=B_n^+\vert_\Upgamma-B^-_n\vert_\Upgamma=\frac{T}{2}\left(\mu_m-\mu_0\right)\left(\bs{\nabla}_s\cdot\msb{C}_s^{-1}(\bs{\nabla}_s(\phi^+\vert_\Upgamma+\phi^-\vert_\Upgamma)\right)+\mathcal{O}(T^2).
\end{split}
\end{align}
Here, $\msb{C}_s=\hat{\msb{I}}\msb{C}$, where $\hat{\msb{I}}=\msb{I}-\bs{n}\otimes\bs{n}$ is the surface projection operator.
\subsection{Kinematics of bubble deformation}
The bubble is considered spherical in its reference configuration, and accordingly, a spherical coordinate system is employed in the present analysis. In the undeformed configuration of the bubble $(r,\theta,\phi)$ represents the radial, meridional and circumferential coordinates, respectively.  Corresponding unit tangent vector triad is represented as $(\bs{e}_r,\bs{e}_\theta,\bs{e}_\phi).$ The position of the deformed bubble surface is described in terms of the mode shapes of a sphere as
\begin{align}\label{e:2.2.1}
    \rho(\theta,t)=R(t)+\sum_{n=2}^{\infty}a_n(t)P_n(\cos{\theta}),
\end{align}
where $P_n$ is Legendre polynomial, $R(t)$ is deformed radius and $a_n(t)$ is amplitude of radial mode shape. The deformed and undeformed mid-surfaces of the encapsulation are,
\begin{align}\label{e:2.2.2}
\begin{split}
\bs{x}&=\rho(\theta,t)\,\bs{e}_r,\\
\bs{X}&=R_0\,\bs{e}_r.
\end{split}
\end{align}
From this the deformation gradient $(\msb{F})$ and right Cauchy stretch tensor $(\msb{C}=\msb{F}^{T}\msb{F})$  can be calculated as
\begin{align}\label{e:2.2.3}
\begin{split}
\msb{F}&=\frac{\partial\bs{x}}{\partial\bs{X}}=\bs{g}_\alpha\otimes\bs{G}^{\alpha}+\lambda_3(\bs{n}\otimes\bs{N}),\\
\msb{C}&=g_{\alpha\beta}(\bs{G}^\alpha\otimes\bs{G}^\beta)+\lambda_3^2(\bs{N}\otimes\bs{N}),
\end{split}
\end{align}
Here, Greek indices $(\alpha,\beta)$ take values $1$ and $2$ correspond to meridional $(\theta)$ and circumferential $(\phi)$ coordinates, respectively. ${\bs g}_\alpha ({\bs G}^\alpha)$ represents the covariant (contra variant) tangent vectors in the deformed (undeformed) membrane mid-surface  with $g_{\alpha\beta}=\bs{g}_\alpha\cdot\bs{g}_\beta$; $\lambda_3$ is the ratio of the deformed thickness to the undeformed thickness of the membrane; $\bs{N}$ and $\bs{n}$ are the normals to the undeformed and deformed mid-surface. The covariant tangent vectors $\bs{g}_1$ and $\bs{g}_2$ are given by
\begin{align}\label{e:2.2.4}
\begin{split}
\bs{g}_1&=\bs{x}_{,\theta}=\rho_{,\theta}\,\bs{e}_r+\rho\,\bs{e}_\theta,\\
\bs{g}_2&=\bs{x}_{,\phi}=\rho\,\sin{\theta}\,\bs{e}_{\phi},\\
\end{split}
\end{align}
and, $\bs{G}^1$ and $\bs{G}^2$ are given by
\begin{align}\label{e:2.2.5}
\bs{G}^1=\frac{\bs{e}_\theta}{R_0},
\bs{G}^2=\frac{\bs{e}_\phi}{R_0\sin{\theta}}.
\end{align}
The resulting right Cauchy-Green deformation tensor $\msb{C}$ tensor as,
\begin{align}\label{e:2.2.6}
\begin{split}
\msb{C}&=\frac{\rho^2+\rho_{,\theta}^2}{R_0^2}(\bs{e}_{\theta}\otimes\bs{e}_{\theta})+\frac{\rho^2}{R_0^2}(\bs{e}_{\phi}\otimes\bs{e}_{\phi})+\lambda_3^2(\bs{e}_r\otimes\bs{e}_r),
\\
&=\lambda_1^2(\bs{e}_{\theta}\otimes\bs{e}_{\theta})+\lambda_2^2(\bs{e}_{\phi}\otimes\bs{e}_{\phi})+\lambda_3^2(\bs{e}_{r}\otimes\bs{e}_{r}).
\end{split}
\end{align}
By comparing coefficients of basis tensor and using incompressibility constraint $\rm{det}~\msb{C}=1$ gives principle stretches $(\lambda_1,\lambda_2,\lambda_3)$ as
\begin{align}\label{e:2.2.7}
\lambda_1=\frac{\sqrt{\rho^2+\rho_{,\theta}^2}}{R_0},\quad
\lambda_2=\frac{\rho}{R_0}\quad\rm{and}\quad
\lambda_3=\frac{1}{\lambda_1\lambda_2}.
\end{align}
Substituting for $\rho$ from~\eqref{e:2.2.1} gives the expression for principle stretches ignoring quadratic terms as 
\begin{align}
\begin{split}
\lambda_1&=\frac{R(t)}{R_0}+\sum_{n=2}^{\infty}\frac{a_n(t)}{R_0}P_n(\cos{\theta})+\mathcal{O}(a_n(t)^2),\\
\lambda_2&=\frac{R(t)}{R_0}+\sum_{n=2}^{\infty}\frac{a_n(t)}{R_0}P_n(\cos{\theta})+\mathcal{O}(a_n(t)^2).
\end{split}
\end{align}
\subsection{Magnetic field induced in the membrane}
For convenience the $\cos\theta$ argument in $P_n(\cos\theta)$ is dropped in following discussion. The outside and inside regions of the undeformed bubble are considered to be $+$ and $-$, where the respective magnetic potential at interface $(\Upgamma)$ is given as
\begin{align}\label{e:2.3.1}
\begin{split}
\phi^+\vert_\Upgamma&=-\frac{\delta A_0}{R_0}-\left(\frac{\delta A_1}{R_0^2}+H_0R_0\right)P_1-\sum_{n=2}^{\infty}\frac{\delta A_nP_n}{R_0^{n+1}},\\
\phi^-\vert_\Upgamma&=-\delta B_0-\left(\delta B_1R_0+H_0R_0\right)P_1-\sum_{n=2}^{\infty}\delta B_nR_0^nP_n.
\end{split}
\end{align}
Also, the corresponding component of the magnetic field in normal direction is
\begin{align}\label{e:2.3.2}
\begin{split}
H_n^+\vert_\Upgamma&=-\frac{\delta A_0}{R_0^2}+\left(H_0-\frac{2\delta A_1}{R_0^3}\right)P_1-\sum_{n=2}^{\infty}(n+1)\frac{\delta A_nP_n}{R_0^{n+2}},\\
H_n^-\vert_\Upgamma&=\left(\delta B_1+H_0\right)P_1+\sum_{n=2}^{\infty}k\delta B_nR_0^{n-1}P_n.
\end{split}
\end{align}
Here, the bookkeeping parameter $\delta$, which does not affect the final results, is introduced to account for the weak magnetic interface, whose presence only slightly perturbs the applied magnetic field. Accordingly, the shape mode amplitudes are rescaled as $a_k=\delta\bar{a}_k$. Using~\eqref{e:15_c}, the corresponding magnetic flux can then be calculated.
\begin{align}\label{e:2.3.3}
B_n=\mu\lambda_1^2\lambda_2^2H_n=\mu\left(\frac{R^4}{R_0^4}+\sum_{k=2}^{\infty}\frac{4R^3}{R_0^3}\delta \bar{a}_nP_n\right)H_n.
\end{align}
From this, the jump in potential and flux becomes,
\begin{align}\label{e:2.3.4}
\begin{split}
[\phi]&=\delta\left(B_0-\frac{A_0}{R_0}\right)+\delta\left(B_1R_0-\frac{A_1}{R_0^2}\right)P_1+\delta\sum_{n=2}^{\infty}\left(B_nR_0^n-\frac{A_n}{R_0^{n+1}}\right)P_n,\\
[B_n]&=\mu_0\frac{R^4}{R_0^4}\left[-\delta\left(\frac{A_0}{R_0^2}\right)-\delta\left(B_1+\frac{2A_1}{R_0^3}\right)-\delta\sum_{n=2}^{\infty}\left(\frac{(n+1)A_n}{R_0^{n+2}}+nB_nR_0^{n-1}\right)\right].
\end{split}
\end{align}
The above equations are truncated to $\mathcal{O}(\delta)$. Substituting these in~\eqref{e:16}, the right hand side terms are truncated to $\mathcal{O}(\delta^2)$ by considering $T\sim\mathcal{O}(\delta^2)$ and $H_0\sim\mathcal{O}(1/\delta)$ which simplifies them as, 
\begin{align}\label{e:2.3.5}
\begin{split}
&[\phi]=\frac{T}{2}\left(\frac{1}{\mu_0}-\frac{1}{\mu_m}\right)(2\mu_0H_0P_1),\\
&[B_n]=-4\frac{T}{2}\left(\mu_m-\mu_0\right)\frac{H_0R_0P_1}{R^2}.
\end{split}
\end{align}
In above equations on right hand side, only $P_1$ term is present which makes all $A_i$ and $B_i$ to be zero to order $\mathcal{O}(\delta^2)$ except $A_1$ and $B_1$ whose equations are
\begin{align}\label{e:2.3.6}
\begin{split}
&\delta B_1-\delta\frac{A_1}{R_0^3}-2q_1\frac{H_0}{R_0}=0,\\
&\delta B_1+2\delta\frac{A_1}{R_0^3}+4q_2\frac{H_0R_0^5}{R^6}=0,
\end{split}
\end{align}
where $q_1=\frac{T}{2}\left(1-\frac{\mu_0}{\mu_m}\right)$ and $q_2=\frac{T}{2}\left(\frac{\mu_m}{\mu_0}-1\right)$. Solving these equations, the coefficients are calculated as,
\begin{align}\label{e:2.3.7}
\begin{split}
\delta A_1&=-\frac{2R_0^2(q_1R^6+2q_2R_0^6)}{3R^6}H_0,\\
\delta B_1&=\,\,\frac{4(q_1R^6-q_2R_0^6)}{3R_0R^6}H_0.
\end{split}
\end{align}
Using the expressions~\eqref{e:2.3.1}, the potential at the top and bottom of the original bubble surface becomes,
\begin{align}\label{e:2.3.8}
\phi^0\lvert_{\Upgamma_2}=\phi^++\frac{T}{2}\frac{\partial\phi^{+}}{\partial r},\quad\quad\quad
\phi^0\lvert_{\Upgamma_1}=\phi^--\frac{T}{2}\frac{\partial\phi^{-}}{\partial r}.
\end{align}
This gives the magnetic field at the mid surface as,
\begin{align}\label{e:2.3.9}
H^0_r=-\frac{\phi^0\lvert_{\Upgamma_2}-\phi^0\lvert_{\Upgamma_1}}{T}=h_1\cos{\theta},\quad\quad\quad
H^0_\theta=-\frac{1}{2R_0}\frac{\partial(\phi^0\lvert_{\Upgamma_2}+\phi^0\lvert_{\Upgamma_1})}{\partial\theta}=h_2\sin{\theta}.
\end{align}
where expressions for $h_1$ and $h_2$ are as given below
\begin{align}\label{e:2.3.10}
\begin{split}
h_1=\left(1-\frac{2q_1}{T}+\frac{4q_1}{3R_0}+\frac{2q_2R_0^5}{3R(t)^6}\right)H_0,\quad
h_2=\left(-1-\frac{q_1}{3R_0}+\frac{4q_2R_0^5}{3R(t)^6}\right)H_0.
\end{split}
\end{align}
\subsection{Equations governing radial and shape modes}
The Lagrangian energy density approach used by~\cite{shaw2006translation} is adopted to model non-spherical oscillations of magnetic bubbles. The expression for the Lagrangian is 
\begin{equation}\label{e:2.4.1}
\mathcal{L}^{*}=\mathcal{C}-\mathcal{T}-\mathcal{K},
\end{equation}
where $(\mathcal{C},\mathcal{T},\mathcal{K})$ corresponds to constraint, potential and kinetic energy respectively, whose expressions are,
\begin{align}
\begin{split}
\mathcal{C}&=-2\pi\rho_f\int_{0}^{2\pi}\psi\rho\frac{\partial\rho}{\partial t}\sin{\theta}\,{\rm d}\theta,\\
\mathcal{T}&=\sigma\lvert S\rvert-V\left[\frac{p_{go}}{1-k}\left(\frac{V_0}{V}\right)^k-p_\infty\right],\\
\mathcal{K}&=\int_{V_f}\frac{1}{2}\rho_{f}\lvert\bs{\nabla}\psi\rvert^2\,{\rm d}V_f,
\end{split}
\end{align}
where $\rho_f$ is the density of the surrounding fluid, $\sigma$ is the surface tension of the bubble, $\psi$ is the velocity potential of the surrounding fluid, $p_{go}$ is the gas pressure inside the bubble in the undeformed state, $p_\infty$ is the pressure at infinity which is sum of ambient and applied pressure, $k$ is the ratio of specific heats of the gas, $S$ is the surface area of the deformed bubble, and $V$ is the volume enclosed by the deformed bubble. The integration over $V_f$ is carried out in the surrounding fluid domain. The velocity potential $\psi$ in the surrounding fluid satisfies the Laplace equation, whose solution in spherical coordinates is
\begin{align}
\psi=-\frac{b_0(t)}{r}+\sum_{n=2}^\infty\frac{b_n(t)}{r^{n+1}}P_n(\cos{\theta}).
\end{align}
The coefficient $b_n$ can be found by following condition from~\cite{shaw2006translation,dash2024nonspherical} 
\begin{align}
\frac{\partial\mathcal{L^*}}{\partial b_n}=0
\end{align}
The presence of magnetic nanoparticles in encapsulation modifies the Lagrangian as
\begin{equation}\label{e:2.4.2}
\mathcal{L}=\mathcal{L}^{*}-\mathcal{U}+\mathcal{M}.
\end{equation}
Here, $\mathcal{U}$ is elastic energy and $\mathcal{M}$ is magnetic energy given by
\begin{subequations}\label{e:2.4.3}
\begin{align}
\mathcal{U}&=T\int_{S_m}\Psi_s(\msb{C})\,{\rm d}S_m=TR_0^2\int_0^{\pi}\Psi_s(\msb{C})\sin{\theta}\,{\rm d}\theta,\label{e:2.4.3_a}\\
\mathcal{M}&=\frac{T\mu_0\chi}{2}\int_{S_m}\msb{C}^{-1}\bs{H}\cdot\bs{H}\,{\rm d}S_m\\&=\frac{T\mu_0\chi R_0^2}{2}\int_{0}^{\pi}\msb{C}^{-1}\bs{H}\cdot\bs{H}\sin{\theta}\,{\rm d}\theta.\label{e:2.4.3_b}
\end{align}
\end{subequations}
Here, $S_m$ denotes the mid-surface of the reference configuration, $\chi$ is the magnetic susceptibility, and $\psi_s(\msb{C})$ represents the hyperelastic strain energy density. The resulting formulation provides the governing equations for the radial and shape modes as follows
\begin{align}\label{e:2.4.4}
\begin{split}
&\frac{{\rm d}}{{\rm d}t}\left(\frac{\partial\mathcal{L}}{\partial\dot{R}}\right)-\frac{\partial\mathcal{L}}{\partial{R}}=0,\\
&\frac{{\rm d}}{{\rm d}t}\left(\frac{\partial\mathcal{L}}{\partial\dot{a}_n}\right)-\frac{\partial\mathcal{L}}{\partial{a_n}}=0.
\end{split}
\end{align}
The neo-Hookean strain energy is considered in our analysis which is,
\begin{align}\label{e:2.4.5}
\Psi_s(\msb{C})=\frac{\mu_s}{2}(\rm{tr(\msb{C})-3})=\frac{\mu_s}{2}(\lambda_1^2+\lambda_2^2+\lambda_3^2-3).
\end{align}
Substituting $\lambda_1$, $\lambda_2$, and $\lambda_3$ from~\eqref{e:2.2.7} into~\eqref{e:2.4.5}, and using the expression for $\bs{H}$ from~\eqref{e:2.3.9} in~\eqref{e:2.4.3_b}, the resulting expressions are expanded up to $\mathcal{O}(a_n^4)$. Integrating the resulting expressions for $\mathcal{U}$ and $\mathcal{M}$ gives the corresponding potential and magnetic energies. Substituting these expressions into~\eqref{e:2.4.4} yields the following governing equations for the radial and non-spherical modes as
\begin{align}\label{e:2.4.6}
\begin{split}
&R\ddot{R}+\frac{3}{2}\dot{R}^2-\frac{1}{\rho_f}\left[p_{go}\left(\frac{R}{R_0}\right)^{3k}-p_{\infty}\right]+\frac{2\sigma}{\rho_fR}+4\eta\frac{\dot{R}}{R}+\frac{2\mu_sT}{\rho_fR}\left[1-\frac{R_0^6}{R^6}\right]\\&+\frac{1}{R}\sum_{n=2}^{\infty}\frac{1}{(2n+1)(n+1)}\left[(n+\frac{3}{2})\dot{a}_n^2+(n+3)a_n\ddot{a}_n
-(n-3)\left(\frac{\ddot{R}}{R}a_n^2\right.\right.\\&\left.\left.+\frac{1}{2}\frac{\dot{R}^2}{R^2}a_n^2+2\frac{\dot{R}}{R}a_n\dot{a}_n\right)\right]+\frac{1}{\rho_fR^2}\left[p_{\infty}-p_{go}(1-3k)\left(\frac{R_0}{R}\right)^{3k}\right]\sum_{n=2}^{\infty}\frac{a_n^2}{2n+1}\\&+\frac{3\mu_sTR_0^6}{\rho_fR^9}\sum_{n=2}^{\infty}\frac{n(n+1)-10}{2n+1}a_n^2-\frac{1}{2\rho_fR^2}\frac{\partial\mathcal{M}}{\partial R}=0,
\end{split}    
\end{align}
\begin{align}\label{e:2.4.7}
\begin{split}
&R\ddot{a}_n+\left(3\ddot{R}+\eta(n+2)(2n+1)\frac{2}{R}\right)\dot{a}_n+(n+1)(n-1)(n+2)\frac{\sigma}{\rho_fR^2}a_n\\&+(n-1)\left((n+2)\eta\frac{2\dot{R}}{R^3}-\ddot{R}\right)a_n
+(n+1)\frac{\mu_sT}{\rho_fR^2}\left(\frac{R_0^6(10-n(n+1))}{R^6}+n^2+n+2\right)a_n\\&+\frac{(2n+1)(n+1)}{4R}
\sum_{i=2}^{\infty}\sum_{j=2}^{\infty}\left[\frac{\ddot{R}}{R}a_ia_jG_{d_{ijn}}+\frac{\dot{R}^2}{R^2}a_ia_jM_{a_{nij}}+\frac{\dot{R}}{R}a_i\dot{a}_jM_{b_{nij}}\right.\\&+\Bigg.a_i\ddot{a}_jM_{c_{nij}}+\dot{a}_i\dot{a}_jM_{d_{nij}}\Bigg]+\frac{(2n+1)(n+1)}{2\rho_fR^2}\left[p_{\infty}-p_{go}\left(\frac{R}{R_0}\right)^{3k}\right]\sum_{i=2}^{\infty}\sum_{j=2}^{\infty}a_ia_jI_{a_{nij}}\\&-\frac{3\mu_sT(2n+1)(n+1)R_0^6}{\rho_fR^9}\sum_{i=2}^{\infty}\sum_{j=2}^{\infty}\left(5I_{a_{nij}}+\frac{1}{2}I_{c_{nij}}+I_{c_{ijn}}\right)a_ia_j\\&-\frac{(n+1)(2n+1)}{2\rho_fR^2}\frac{\partial\mathcal{M}}{\partial a_n}=0,
\end{split}
\end{align}
where $\eta$ is viscosity of surrounding fluid and viscous terms are added from~\cite{shaw2009stability} and truncated to order $\mathcal{O}(a_n^2)$. The various coefficients $I, G, M$ can be found in~\cite{shaw2006translation}. Also the expression for $\mathcal{M}$ in terms of radius and mode amplitudes $a_n$ is given in Appendix~\ref{A:1}.
\subsection{Natural frequency calculation}
The natural frequencies of the radial and shape modes can be obtained by substituting $R(t)=R_0+x(t)$ into~\eqref{e:2.4.6} and~\eqref{e:2.4.7}, respectively, and linearizing the resulting equations with respect to $x(t)$ and $a_n(t)$. By retaining the linear terms and extracting the corresponding coefficients of $x(t)$ and $a_n(t)$, the following expressions for the natural frequencies are obtained.
\begin{equation}\label{e:2.5.2}
\omega_r^2=3k\frac{p_{go}}{\rho_fR_0^2}+\frac{12\mu_sT}{\rho_fR_0^3}-\frac{2\sigma}{\rho_fR_0^3}+\frac{8H_0^2\mu_0\chi q_1}{3\rho_fR_0^3},
\end{equation}
\begin{align}\label{e:2.5.1}
\begin{split}
\omega_k^2&=(n+1)(n-1)(n+2)\frac{\sigma}{\rho_fR_0^3}+(n+1)\frac{12T\mu_s}{\rho_fR_0^3}\\&-\mu_0\chi T\frac{(n+1)(2n+1)(28n^3+40n^2-3n-9)h_1^{2}}{(8n^3+12n^2-2n-3)\rho_fR_0^3}\\&+\mu_0\chi T\frac{(n+1)(2n+1)(4n^3+6n^2-13n+6)h_2^2}{(4n^2-1)\rho_fR_0^3},
\end{split}
\end{align}
where $h_1$ and $h_2$ are evaluated at $R(t)=R_0$.
\section{Results and discussion}\label{s:3}
\subsection{Induced magnetic field}
\begin{figure}
    \centering
    \includegraphics[width=1\linewidth]{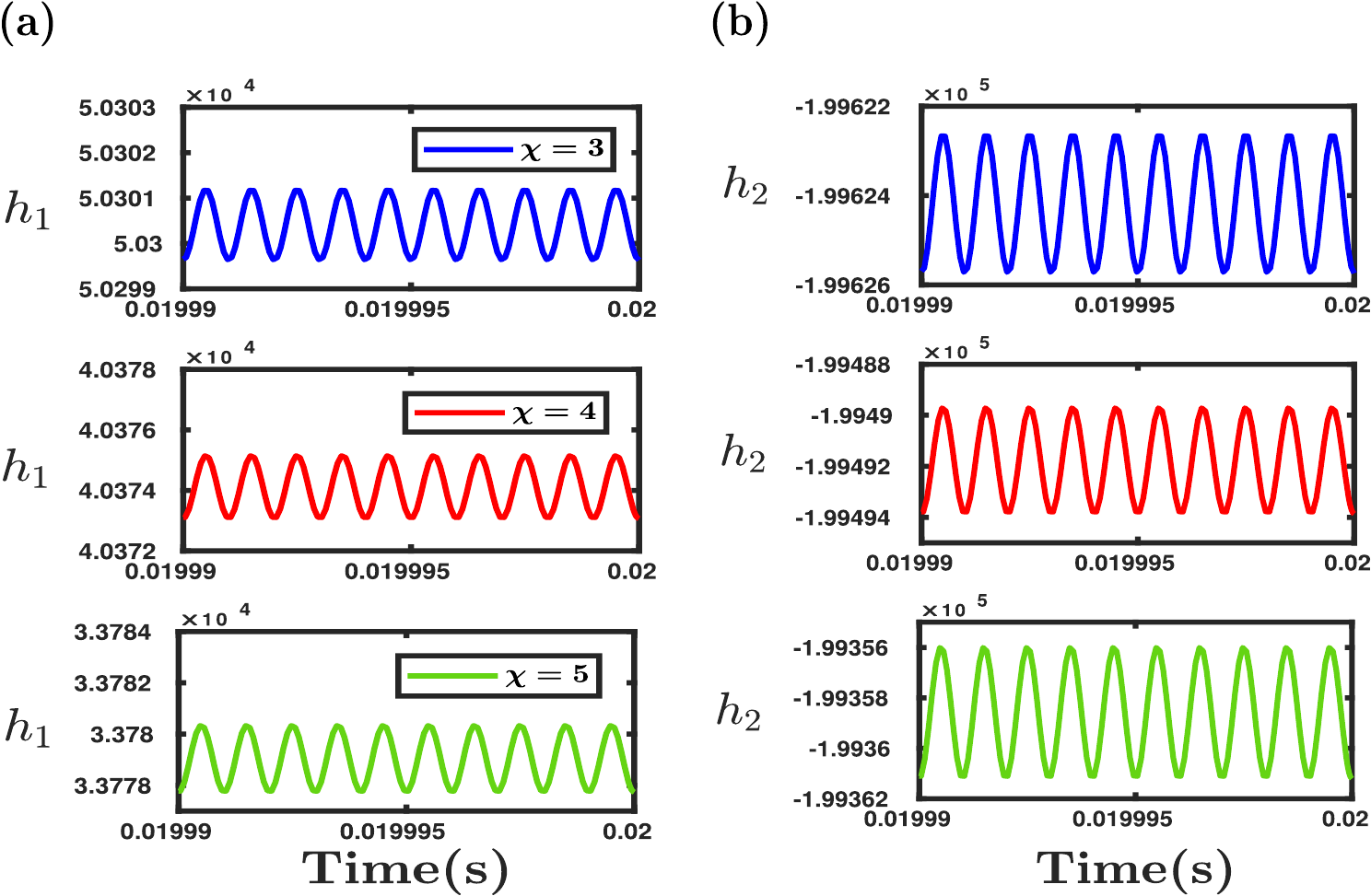}
    \caption{Time series of $h_1$ and $h_2$ for a bubble of initial radius $R_0=20\,\mu\rm{m}$, surface shear modulus $\mu_ST=0.1\,\rm{Nm}^{-1}$ under driving frequency $f_d=1\,\rm{MHz}$ and driving pressure $p_A=100\,\rm{kPa}$ for different values of susceptibility $\chi$.}
    \label{fig:2}
\end{figure}
This section presents the magnetic field induced in the reference configuration, known as the Lagrangian magnetic field $\bs{H}$. The pressure and magnetic field are applied as follows,
\begin{align}\label{e:5.1}
p_{\infty}=p_{0}+p_a\cos{2\pi f_dt},\quad\quad H=H_0.
\end{align}
Here, $p_0=10^3\,\rm{kPa}$ is the ambient pressure, $p_a$ is the amplitude of the time-varying pressure, $H_0$ is the applied magnetic field at infinity, and $f_d$ is the driving frequency. As the amplitudes of the normal and tangential components of the induced magnetic field, $h_1$ and $h_2$, vary with the deformed radius, the bubble is excited with a driving frequency of $f_d=1\,\rm{MHz}$, $p_a=100\,\rm{kA}$ and $H_0=200\,\rm{kA/m}$ for a bubble of initial radius $R_0=20\,\mu\rm{m}$.
\begin{figure}
    \centering
    \includegraphics[width=1\linewidth]{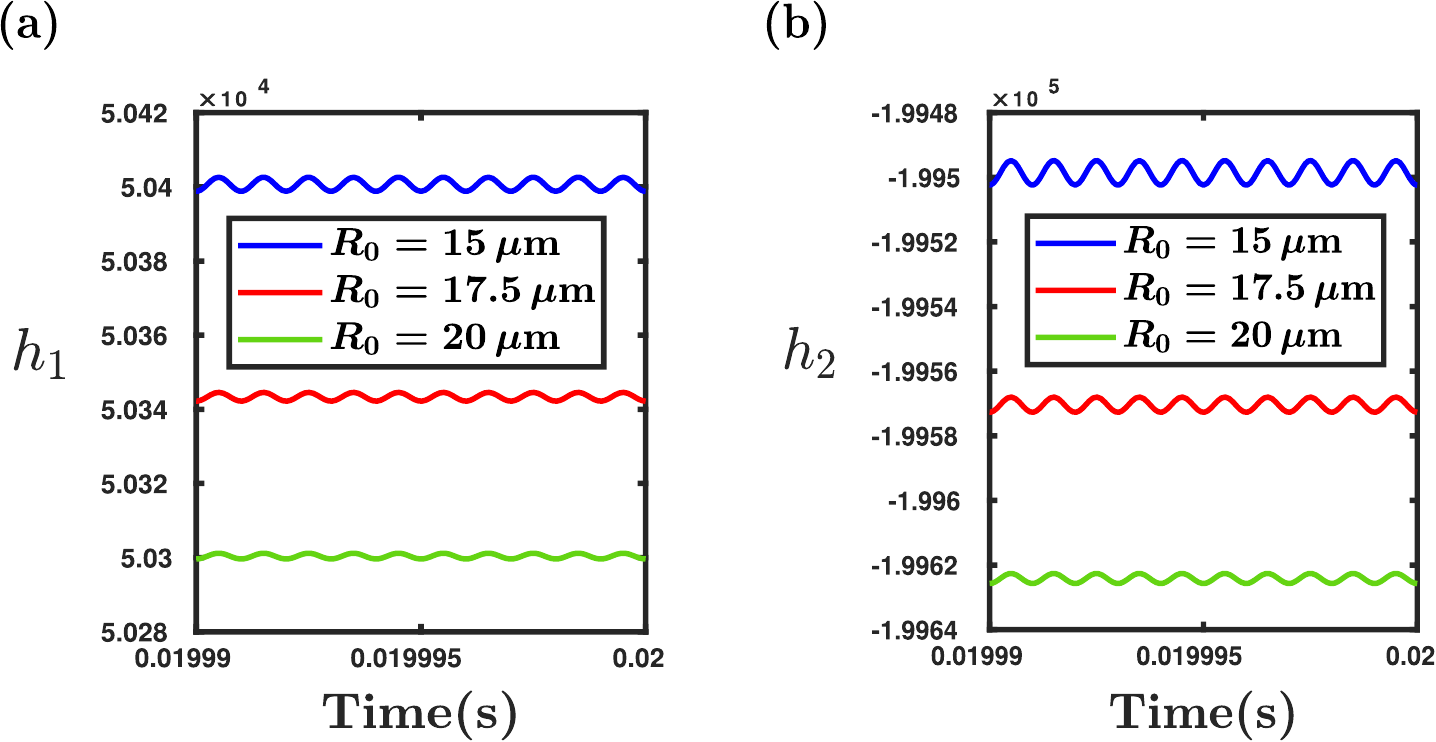}
    \caption{Time series of $h_1$ and $h_2$ for a bubble of susceptibility $\chi=3$, surface shear modulus $\mu_ST=0.1\,\rm{Nm}^{-1}$ under driving frequency $f_d=1\,\rm{MHz}$ and driving pressure $p_A=100\,\rm{kPa}$ for different values of initial radius $R_0$.}
    \label{fig:3}
\end{figure}
It can be observed from $\chi=3$ case of figure~\ref{fig:2} that the normal component of the induced Lagrangian magnetic field is reduced significantly, whereas the tangential component is only slightly reduced. Increasing the size of bubble slightly increases the value of the induced magnetic field, as induced magnetic field is inversely proportional to the undeformed bubble radius. Also, increasing the susceptibility of the encapsulation significantly reduces $h_1$ and slightly increases $h_2$, which follows from~\eqref{e:2.3.10}. These behavior can be seen from figure~\ref{fig:2} and~\ref{fig:3}.
\subsection{Influence of applied magnetic field on natural frequency and stability diagram}
\begin{figure}
    \centering
    \includegraphics[width=1\linewidth]{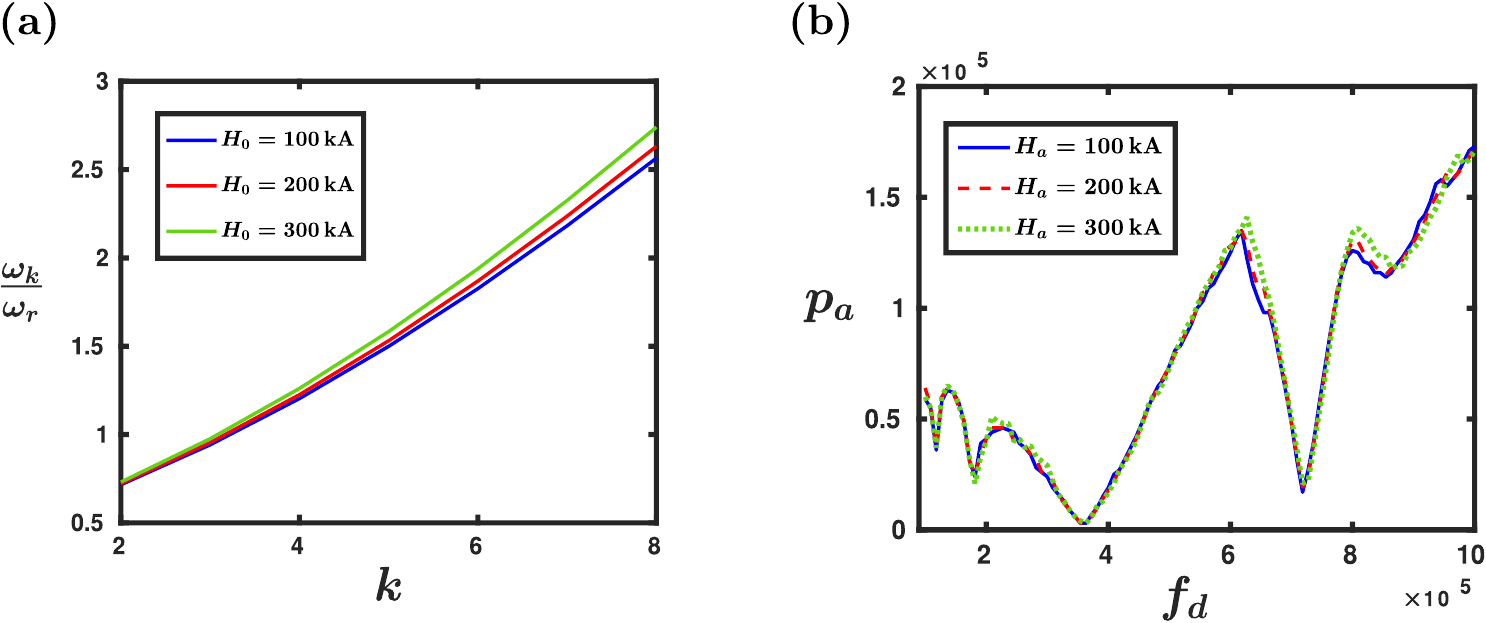}
    \caption{(a) Variation of natural frequency with mode number $\omega_k/\omega_r\,{\rm vs}\,k$, (b) Critical pressure vs driving frequency in the $p_a\,{\rm vs}\,f_d$ plane for a bubble of initial radius $R_0=10\,\mu\rm{m}$, susceptibility $\chi=3$ and surface shear modulus $\mu_sT=0.1\,\rm{Nm}^{-1}$ shown for various values of applied magnetic field.} 
    \label{fig:5}
\end{figure}
In the previous section it is observed that $\lvert h_2\rvert>\lvert h_1\rvert$ and from~\eqref{e:2.5.1} $\omega_k^2\propto h_2^2-h_1^2$. So, increasing the magnetic field increases the natural frequency of the mode shapes. Also, from figure~\ref{fig:5}a, as the mode number is increased, the influence of the applied magnetic field becomes more significant. In this work, the exponential study is considered which is the critical value of applied pressure $p_a$ at which the stable sinusoidal oscillations blow-up exponentially.  The simulations are conducted for a bubble of initial radius $R_0=10\,\mu\rm{m}$, surface shear modulus $\mu_sT=0.1\,\rm{Nm}^{-1}$ and susceptibility $\chi=3$. Figure~\ref{fig:5}b suggests that increasing the applied magnetic field slightly shifts the stability curves to the right and slightly increases the stability. This is because linearized version of~\eqref{e:2.4.7} has structure similar to Mathieu's equation $\ddot{\beta}+(\Delta+\epsilon\cos{\omega_dt})\beta=0$ where $\Delta$ increases as $H_0$ is increased and $\epsilon$ is influenced by $p_a$. Increasing $\Delta$ requires a higher value of $p_a$ to make the system unstable.
\subsection{Influence of magnetic field on frequency response and finite amplitude oscillations}
\begin{figure}
    \centering
    \includegraphics[width=1\linewidth]{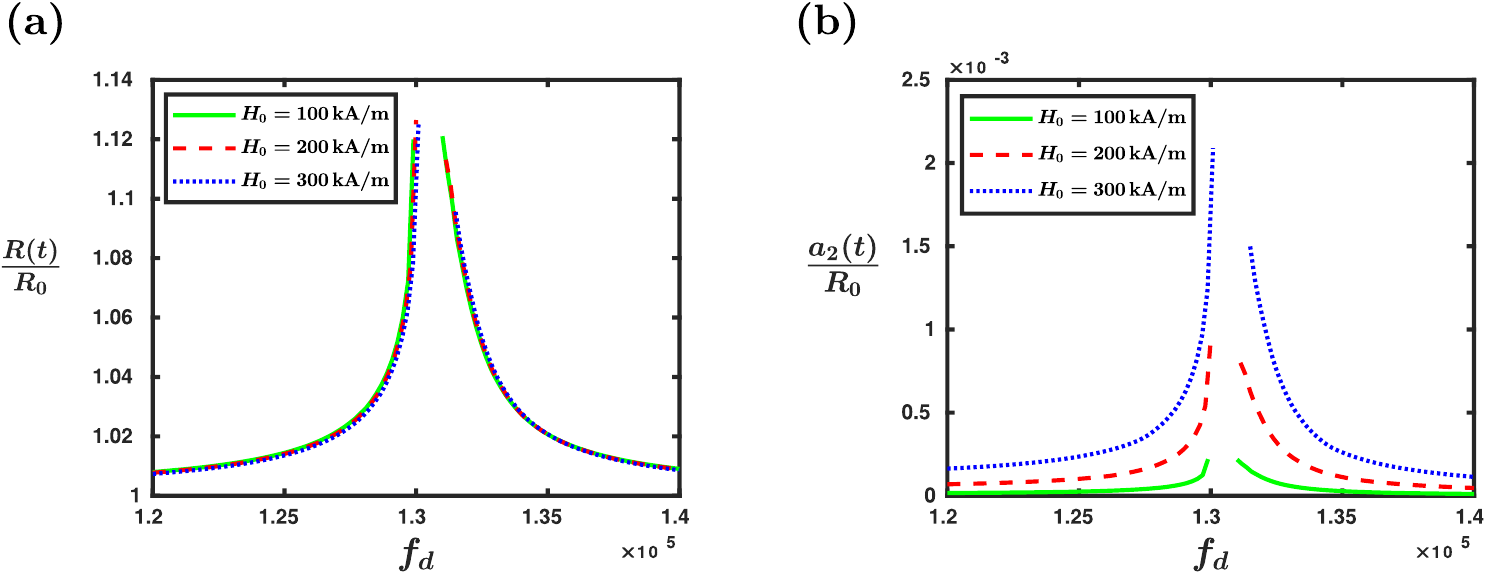}
    \caption{Frequency response of radial and $a_2$ mode for a bubble of undeformed radius $R_0=25\,\mu\rm{m}$, susceptibility $\chi=3$ and surface shear modulus $\mu_sT=0.1\,\rm{Nm}^{-1}$ at different values of applied magnetic field.}
    \label{fig:6}
\end{figure}

As the model considers quadratic shape mode interaction, sustained finite amplitude shape oscillations can be observed when bubble is excited near the resonance frequency. First, the bubble is excited near primary resonance of radial mode. A bubble of initial radius $R_0=25\,\mu\rm{m}$, susceptibility $\chi=3$, surface  shear modulus $\mu_sT=0.1\,\rm{Nm}^{-1}$ is considered in our analysis. From~\eqref{e:2.5.2}, increasing the applied magnetic field slightly increases the radial resonance frequency, which shifts the radial frequency response curves to the right as shown in figure~\ref{fig:6}a. Also, the $a_2$ mode is directly forced by the magnetic field and radial oscillations. So, increasing the magnetic field near the resonant region of radial oscillation significantly increases the $a_2$ oscillation amplitude figure~\ref{fig:6}b.
\begin{figure}
    \centering
    \includegraphics[width=1\linewidth]{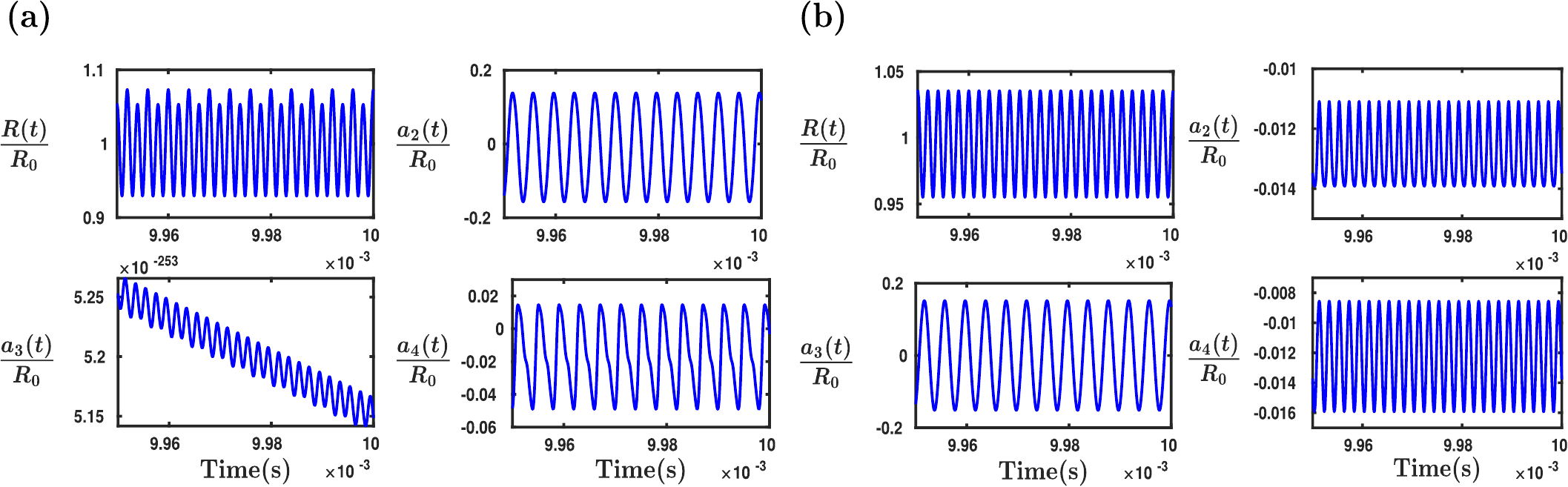}
    \caption{Time series for radial and first three shape modes for bubble under driving pressure frequency of $0.5\,\rm{MHz}$, susceptibility $\chi=3$, surface shear modulus $\mu_sT=0.1\,\rm{Nm}^{-1}$ and initial radius (a) $R_0=12.5\,\mu\rm{m}$ and (b) $R_0=15\,\mu\rm{m}$.}
    \label{fig:7}
\end{figure}
\begin{figure}
    \centering
    \includegraphics[width=1\linewidth]{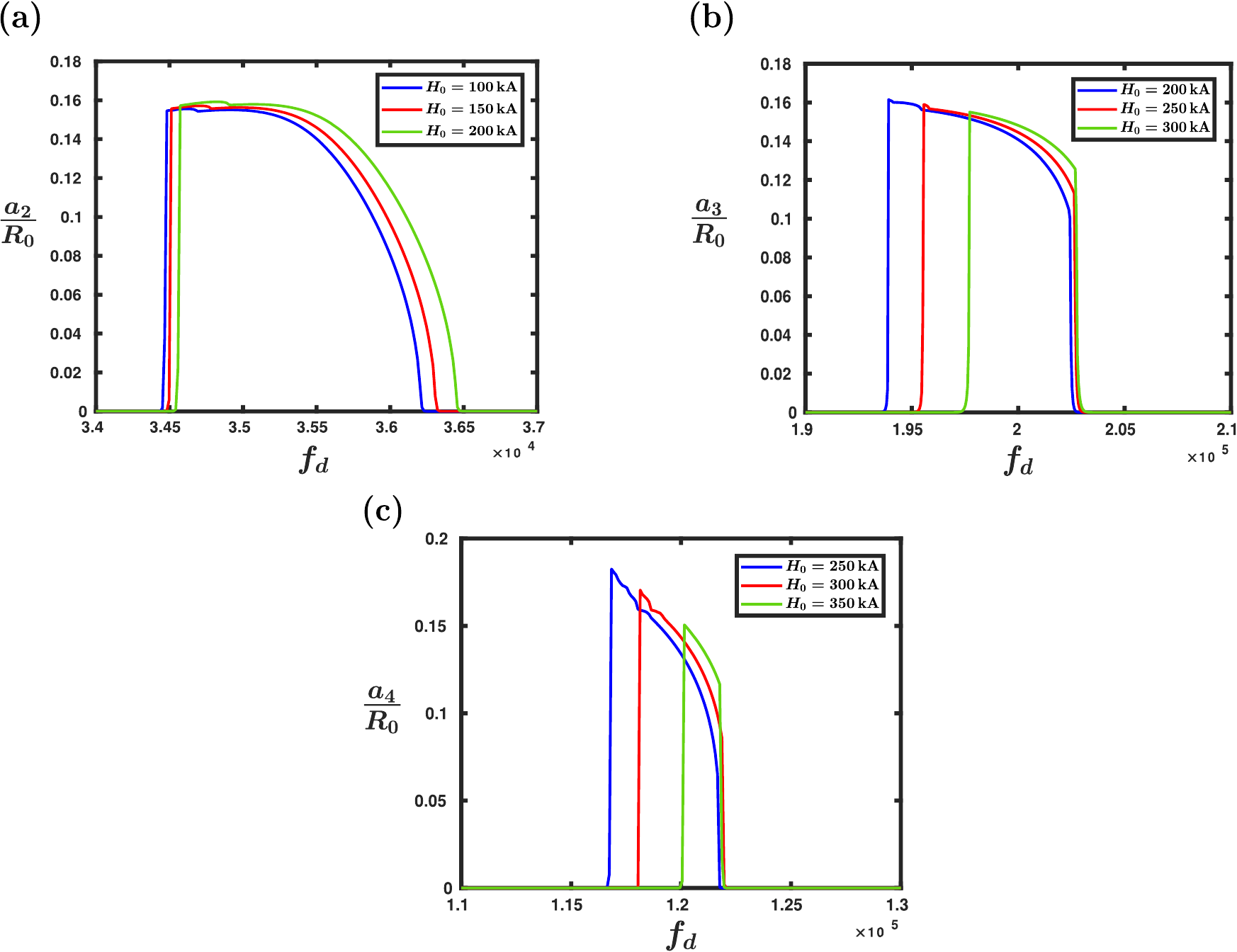}
    \caption{Frequency response at different values of applied magnetic field for a bubble of susceptibility $\chi=3$, surface shear modulus $\mu_sT=0.1\,\rm{Nm}^{-1}$ and initial radius (a) $R_0=72\,\mu\rm{m}$ excited at $p_a=0.4\,\rm{kPa}$, (b) $R_0=27\,\mu\rm{m}$ excited at $p_a=20\,\rm{kPa}$ and (c) $R_0=44.8\,\mu\rm{m}$ excited at $p_a=12.8\,\rm{kPa}$.}
    \label{fig:8}
\end{figure}
\begin{figure}
    \centering
    \includegraphics[width=\linewidth]{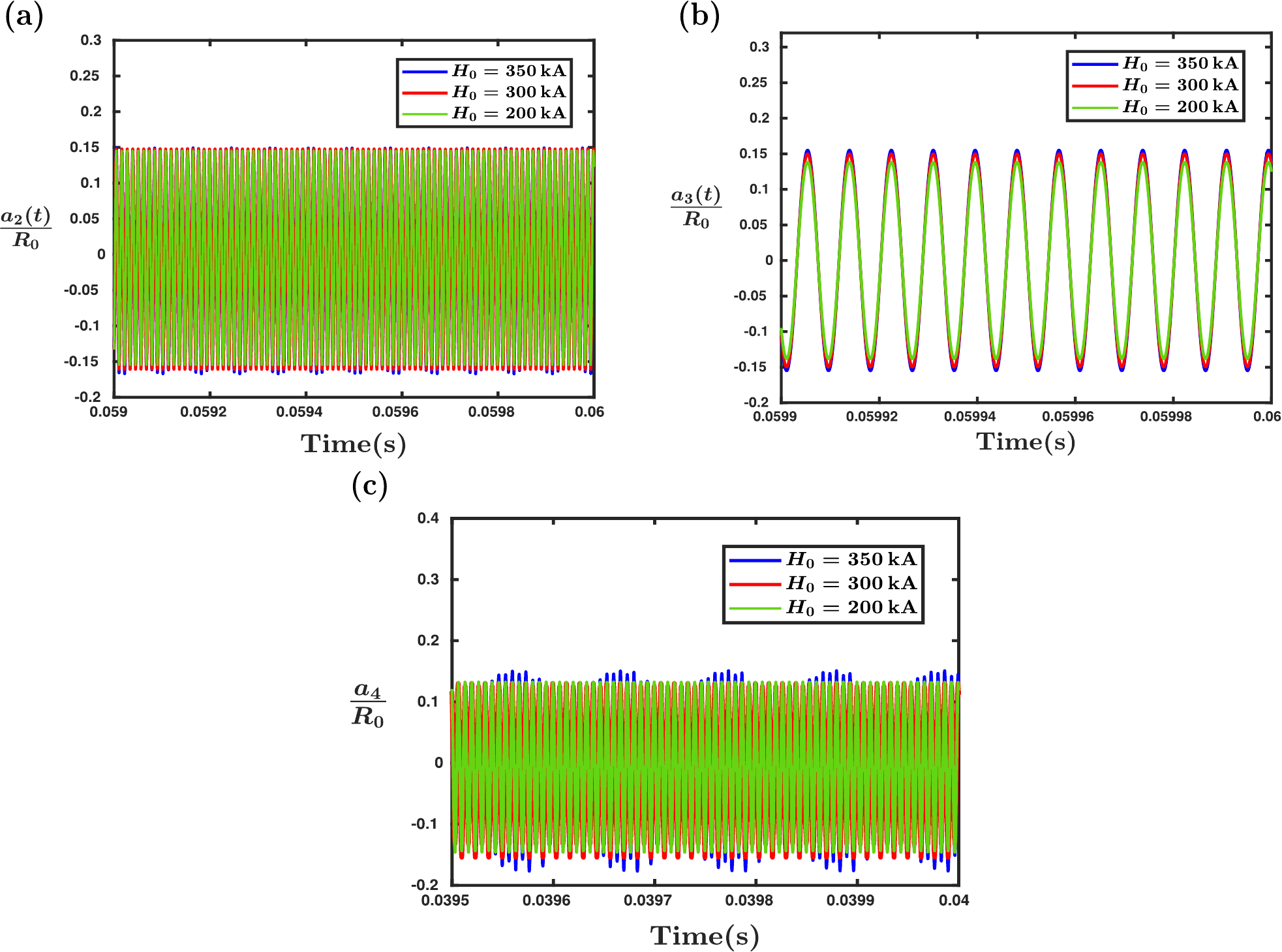}
    \caption{Time series of second, third and fourth mode for a bubble of initial radius $R_0=25\,\mu\rm{m}$, susceptibility $\chi=3$ and surface shear modulus $\mu_sT=0.1\,\rm{Nm}^{-1}$ at (a) $f_d=175\,\rm{kHz}$,$\,p_a=15\,\rm{kPa}$, (b) $f_d=234\,\rm{kHz}$,$\,p_a=30\,\rm{kPa}$ and (c) $f_d=300\,\rm{kHz}$,$\,p_a=48\,\rm{kPa}.$}
    \label{fig:9}
\end{figure}
The shape modes are parametrically excited by the radial mode. In this study, only the first parametric resonance is considered that is, $\omega_d\approx2\omega_k$ for which a particular mode undergoes higher amplitude of finite oscillations compared to other modes if a threshold pressure is reached. The driving frequency and applied magnetic field considered is $f_d=0.5\,\rm{MHz}$, $H_0=200\,\rm{kA}$. Then, for a bubble of radius $R_0=12.5\,\mu\rm{m}$, the first three shape mode natural frequencies are $2f_k=\left[0.487,0.6445,0.8271\right]\,\rm{MHz}$ . If $p_a/p_0=0.7$, then sustained finite-amplitude oscillation for the second mode is seen in figure~\ref{fig:7}a. Similarly, for a bubble of radius $R_0=15\,\mu\rm{m}$, the first three shape mode natural frequencies are $2f_k=\left[0.3705,0.4903,0.6292\right]\,\rm{MHz}$, and the first parametric resonance of the third mode is near the driving frequency.  If $p_a/p_0=0.7$, then sustained finite-amplitude oscillation  
for the third mode is observed as shown in figure~\ref{fig:7}b. The threshold pressure is in general not the same and in this example, they are the same.
\begin{figure}
    \centering
    \includegraphics[width=1\linewidth]{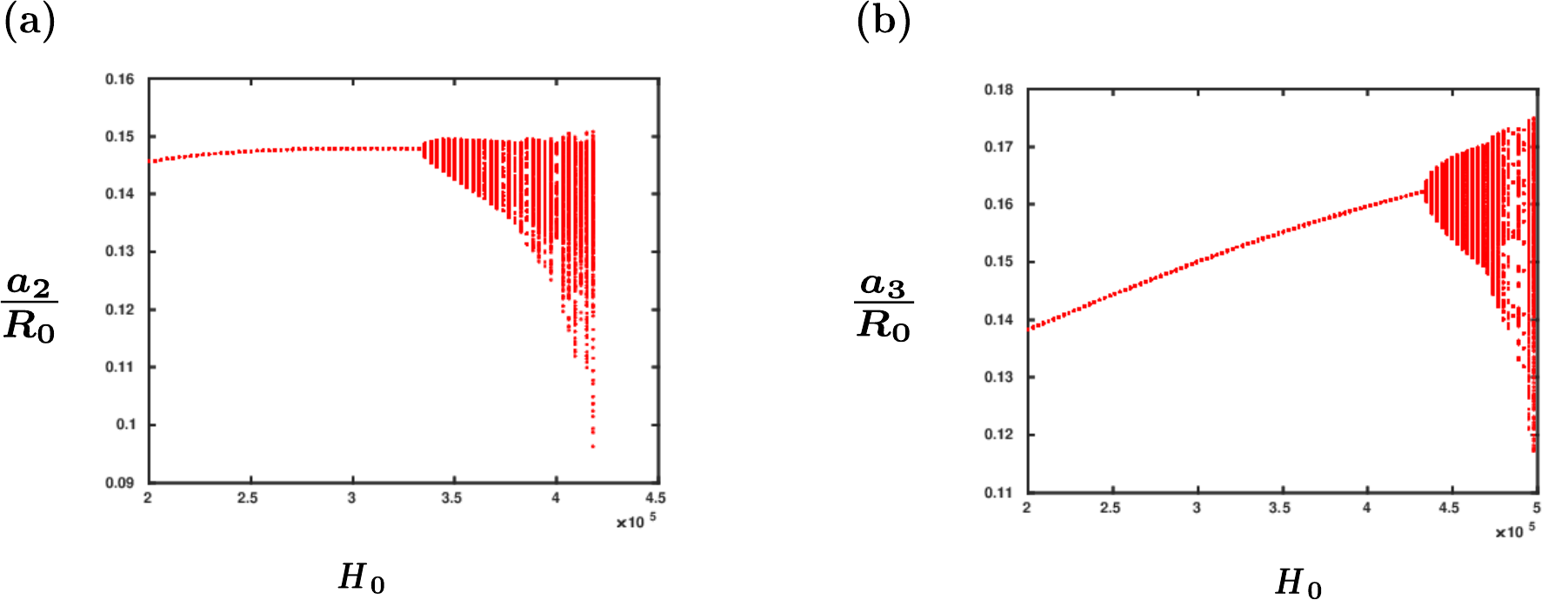}
    \caption{Bifurcation diagram of (a) $a_2$ mode and (b) $a_3$ mode with $H_0$.}
    \label{fig:10}
\end{figure}
\begin{figure}
    \centering
    \includegraphics[width=\linewidth]{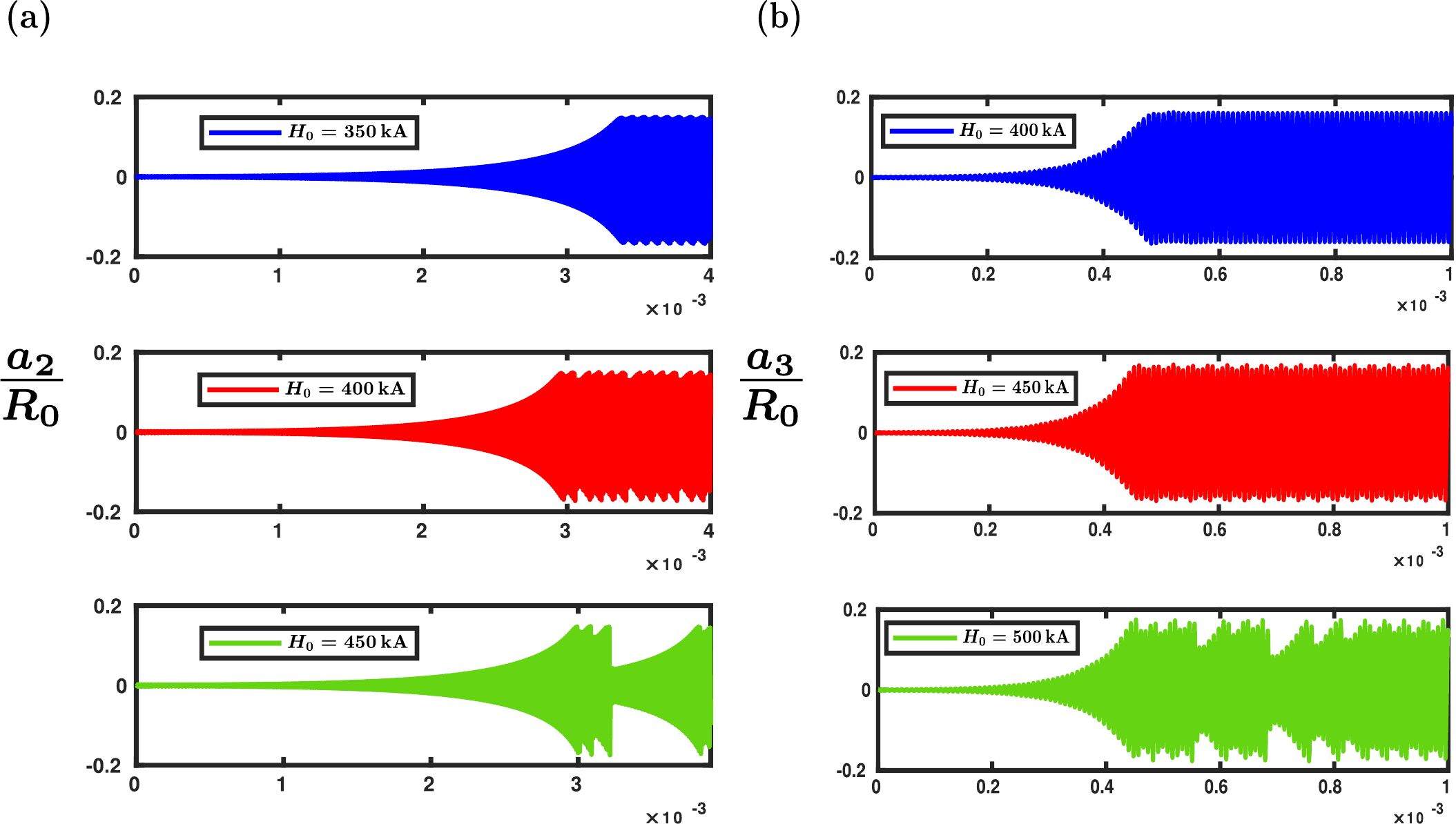}
    \caption{Transition from stable periodic oscillation to highly chaotic oscillation and exponentially blowing up with increase in applied magnetic field.}
    \label{fig:11}
\end{figure}

To understand the influence of the magnetic field on finite amplitude oscillation, the frequency response curves are shown in figure~\ref{fig:8} for various values of applied magnetic field. A bubble of initial radius $R_0=72\,\mu\rm{m}$, susceptibility $\chi=3$, surface shear modulus $\mu_sT=0.1\,\rm{Nm}^{-1}$is excited at $p_a=0.4\,\rm{kPa}$ in the range of driving frequency $f_d=0.34-0.37\,\rm{MHz}$ which is around twice the natural frequency of second mode $2f_2\approx35\,\rm{kHz}$. It can be observed from figure~\ref{fig:8}a that as the applied magnetic field is increased from $H_0=100\,\rm{kA}$ to $H_0=200\,\rm{kA}$, the frequency response curves get shifted to the right, which is consistent with frequency plot in figure~\ref{fig:5}a where the natural frequency increases as the applied magnetic field is increased. Similar pattern is also observed in the $a_3$ mode, shown in figure~\ref{fig:8}b, for a bubble of radius $R_0=27\,\mu\rm{m}$ over the range of magnetic field $H_0=200,\rm{kA}$ to $H_0=300\,\rm{kA}$. Similarly, for the $a_4$ mode, shown in figure~\ref{fig:8}c, the same pattern is observed for a bubble of radius $R_0=44.8\,\mu\rm{m}$ over the range of magnetic field $H_0=250\,\rm{kA}$ to $H_0=350\,\rm{kA}$, when excited around the natural frequencies of the third and fourth modes, $2f_3\approx200\,\rm{kHz}$ and $2f_4\approx120\,\rm{kHz}$, respectively. The range of magnetic field strengths and bubble radii is selected such that no unstable response is observed within the subharmonic frequency range. The influence of the applied magnetic field on the time-series response is not uniform and depends on the applied magnetic field, applied pressure, and driving frequency. 

Depending on various bubble parameters and forcing, the oscillation amplitude can either increase or decrease with increasing magnetic field. However, in the region where finite-amplitude oscillations overlap, an increase in the applied magnetic field generally enhances the maximum amplitude of the mode oscillations. To understand the time series behavior of mode shapes in the overlapping region with an increase in the applied magnetic field, a bubble of radius $R_0=25\,\mu\rm{m}$ is excited near its first parametric resonance for the second, third, and fourth modes under different values of $p_a$, as different modes require different threshold pressures to exhibit finite amplitude oscillations. The driving frequency of the $a_2$ mode where finite-amplitude oscillations overlap in the near-resonant region is $f_d\approx175\,\rm{kHz}$. For the second mode, the threshold pressure is $p_a=15\,\rm{kPa}$, whose timeseries is shown in figure~\ref{fig:9}a. For the third and fourth mode near resonant driving frequency and threshold pressure are $f_d=234\,\rm{kHz}$ at $p_a=30\,\rm{kPa}$ and $f_d=300\,\rm{kHz}$ at $\,p_a=48\,\rm{kPa}$, whose time series are given in figures~\ref {fig:9}b and~\ref{fig:9}c. It can be observed from these plots that, within the overlapping region, increasing the applied magnetic field increases the maximum amplitude of the oscillations and significantly influences their phase and period. Furthermore, the finite-amplitude oscillations become progressively more complex and chaotic as the applied magnetic field is increased. To investigate this behavior, a bifurcation diagram is constructed by taking $H_0$ as the bifurcation parameter and tracking the peak amplitudes of the second and third modes. As shown in figure~\ref{fig:10}a and~\ref{fig:10}b, increasing the applied magnetic field causes the oscillations to undergo a period-doubling sequence, transitioning from period-two to period-four oscillations, followed by further period doublings and the eventual onset of chaotic behavior figure~\ref{fig:11}. With a further increase in the applied magnetic field, the oscillation amplitude  grows rapidly, eventually leading to the failure of the bubble.
\section{Summary and conclusions}\label{s:4}
This study presents a modified theoretical framework for calculating the induced magnetic field in a thin, deformable magneto-elastic membrane. Using this framework, the leading-order terms of the induced magnetic field are obtained. The resulting strain and magnetic energies are incorporated into the potential energy of the Lagrangian formulation to derive the governing equations for the radial and shape modes, including quadratic interactions. The natural frequencies of the shape modes are subsequently obtained by linearizing the mode-shape equations and are observed to increase with increasing applied magnetic field. The exponential stability of the magnetic bubble also increases with increasing applied magnetic field. When the bubble is excited near the primary radial frequency, the applied magnetic field produces only a slight shift in the radial frequency response, while significantly increasing the amplitude of $a_2$ oscillations. For excitation near the first parametric resonance, the frequency response shifts towards higher frequencies with increasing applied magnetic field, consistent with the increase in the natural frequency of the shape modes. Furthermore, when the bubble undergoes finite-amplitude oscillations, the applied magnetic field has a significant influence on the phase and period of the oscillations. Bifurcation analysis, with the applied magnetic field as the bifurcation parameter, shows that the bubble dynamics become progressively more complex and chaotic as the magnetic field is increased, eventually leading to exponential blow-up.

Although the present study successfully demonstrates the influence of the applied magnetic field on finite-amplitude oscillations and frequency response, the model does not account for tangential shape deformations that are present in encapsulated bubbles. Developing a model that incorporates these deformations is challenging, as it requires solving the governing equations in their strong form. Accounting for quadratic shape-mode interactions within the strong-form formulation introduces additional complexity. Furthermore, the calculation of magnetic forces in the strong form requires the gradient of the magnetic field. However, this gradient vanishes when the jump in magnetic potential and magnetic flux is truncated at order $\mathcal{O}(T^2)$, necessitating the inclusion of higher-order terms.

In several cases, the frequency-response analysis reveals a range of driving frequencies over which the system becomes unstable. For such frequency ranges, it is difficult to determine whether the system possesses an unstable branch of the frequency response or whether no periodic solutions exist, owing to the strong nonlinearities in the governing equations. From simulations it was also observed that truncating the radial terms in the mode-shape equations at quadratic order does not adequately approximate the original system, particularly in the resonant regions. The truncated system requires significantly higher applied pressures to trigger finite-amplitude oscillations compared to the original system. Therefore, a more accurate representation of the system would require retaining cubic interactions in the governing equations. Developing the model up to cubic order and subsequently seeking appropriate simplifications would provide a more reliable framework for analyzing the nonlinear dynamics and frequency response of magnetic bubbles.
 
Despite these limitations, the present work provides a foundation for calculating the induced magnetic field in thin encapsulated bubbles subjected to a uniform magnetic field using imperfect interface theory and for incorporating the resulting magnetic energy into the Lagrangian formulation to obtain the shape-mode equations. This framework provides a basis for further development of more comprehensive models that can account for additional nonlinear interactions and tangential deformations.\\

\justify{}
\textbf{Acknowledgement.} AKB would like to thank Shaik Nadeem Karim for useful discussions. AKB and GT acknowledge the Ministry of Education (MoE), Government of India, for providing research support.\\

\justify{}
\textbf{Funding.} This work was supported by the Science and Engineering Research Board, Department of Science and Technology, Government of India, under Project No. CRG/2022/005775.\\
\justify{
\textbf{Declaration of Interests.} The authors report no conflict of interest.\\
}
\appendix
\section{Expression for magnetic energy $\mathcal{M}$}\label{A:1}
\begin{align}
\begin{split}
\mathcal{M}&=\frac{\mu_0\chi R^4}{3R_0^2}h_1^2+\frac{2\mu_0\chi R_0^4}{3R^2}h_2^2+\frac{8\mu_0\chi R^3}{15R_0^2}h_1^2a_2+\frac{4\mu_0\chi R_0^4}{15R_0^2}h_2^2a_2+\frac{\mu_0\chi R^2}{R_0^4}h_1^2\sum_{i=2}^{\infty}\sum_{j=2}^{\infty}a_ia_jH^a_{ij}\\
&+\frac{6\mu_0\chi R^2} {R_0^4}h_1^2\sum_{i=2}^{\infty}\sum_{j=2}^{\infty}a_ia_jH^b_{ij}+\frac{3\mu_0\chi R_0^2}{R^4}h_2^2\sum_{i=2}^{\infty}\sum_{j=2}^{\infty}a_ia_jH^c_{ij}-\frac{\mu_0\chi R_0^4}{R^4}h_2^2\sum_{i=2}^{\infty}\sum_{j=2}^{\infty}a_ia_jH^d_{ij}\\
&+\frac{2\mu_0\chi R}{R_0^4} h_1^2\sum_{i=2}^{\infty}\sum_{j=2}^{\infty}\sum_{k=2}^{\infty}a_ia_ja_k\mathcal{H}^a_{ijk}+\frac{4\mu_0\chi R}{R_0^4} h_1^2\sum_{i=2}^{\infty}\sum_{j=2}^{\infty}\sum_{k=2}^{\infty}a_ia_ja_k\mathcal{H}^b_{ijk}\\
&+\frac{4\mu_0\chi R_0^2}{R^5} h_2^2\sum_{i=2}^{\infty}\sum_{j=2}^{\infty}\sum_{k=2}^{\infty}a_ia_ja_k\mathcal{H}^c_{ijk}-\frac{4\mu_0\chi R_0^2}{R^5} h_2^2\sum_{i=2}^{\infty}\sum_{j=2}^{\infty}\sum_{k=2}^{\infty}a_ia_ja_k\mathcal{H}^d_{ijk}
\end{split}
\end{align}
\begin{align}
\begin{split}
H^a_{ij}&=\int_{-1}^{1}x^2P^1_i(x)P^1_j(x)dx\quad\quad\quad H^b_{ij}=\int_{-1}^{1}x^2P_i(x)P_j(x)dx\\
H^c_{ij}&=\int_{-1}^{1}(1-x^2)P_i(x)P_j(x)dx\quad H^d_{ij}=\int_{-1}^{1}(1-x^2)P^1_iP^1_jdx
\end{split}
\end{align}
\begin{align}
\begin{split}
\mathcal{H}^a_{ijk}&=\int_{-1}^{1}x^2P_i(x)P^1_j(x)P^1_k(x)dx\quad\quad\quad\,\,\,\mathcal{H}^b_{ijk}=\int_{-1}^{1}x^2P_i(x)P_j(x)P_k(x)dx\\
\mathcal{H}^c_{ijk}&=\int_{-1}^{1}(1-x^2)P_i(x)P^1_j(x)P^1_k(x)dx\quad\mathcal{H}^d_{ijk}=\int_{-1}^{1}(1-x^2)P_i(x)P_j(x)P_k(x)dx
\end{split}
\end{align}
\bibliographystyle{jfm}
\bibliography{references}

\end{document}